# Models of CT dose profiles in Banach space; with applications to CT Dosimetry

Victor J Weir

Baylor Scott and White Healthcare system

Dallas TX, 75246

**ABSTRACT**

This paper consists of two parts. In the first part, the scatter components of computed tomography (CT) dose profiles are modeled using various functions including the solution to Riccati's differential equation. These scatter functions are combined with primary components such as the trapezoidal function and a constructed function that uses the analytic continuation of the Heaviside step function. A mathematical theory is developed in a Banach space. The modeled function is used to accurately fit data from the O-arm cone beam CT imaging system. In a second part of the paper, an approach to CT dosimetry is developed that shows that the result obtained from the use of a pencil shaped ion chamber is equivalent to that from a farmer type chamber. This result is verified by presenting some preliminary experimental data measured in a 64 slice Siemens Sensation scanner.

# INTRODUCTION

Over the past few years, the issue of dose in computed tomography (CT) has received a lot of attention. This is partly due to the fact that utilization of CT scanning continues to grow. Typically, the computed tomographic dose index (CTDI)[1, 2], the dose to a uniform polymethyl metacrylate (PMMA) phantom is used to estimate patient dose when patient size corrections are made. CTDI is associated with a homogeneous 32 cm body or 16cm head phantom. It was defined as a single axial exposure to a 100mm pencil shaped ion chamber inserted into a 15cm long body or head phantom. This definition is accurate for narrow single detectors where the entire dose profile, with scatter tails, is within the collection region of the pencil chamber. Beside this definition of the dose index, there were also mathematical descriptions of the dose profile produced by a CT beam in a phantom. These involved the use of a double Gaussian function [2], of Gaussian and Lorentzian functions [3, 4], purely Gaussian function [5], and an exponential function[6]. Furthermore, a mathematical model that described the scatter contribution to the beam profile as a two-dimensional convolution between a primary function and an unknown blurring function was developed by Gagne in 1989 [3]. In that paper, the primary function was modeled as a rectangular function $\prod(\frac{z}{w})$ of unit height and width $w$. As beam collimations got wider due to technological advancements, scatter profiles became wider and scatter tails got longer. This has an impact on patient dosimetry considerations due to the mismatch of the shape and width of the radiation dose profile, as well as long scatter tails falling outside of the extent of the pencil chamber used for measuring CTDI. It was soon realized that the CTDI measured for this wider beams using the pencil chamber would underestimate the actual dose. To help address these issues for wider multi detector scanners with wider dose profile coverage, recent proposals have focused on redefining the CTDI by using a point chamber and a longer body phantom [7, 8]. This approach depends on the theory that as the beam collimation increases, and a longer phantom of greater than 300 mm is used, there is an approach to equilibrium so that a point chamber should be adequate to capture peak exposure from the dose profile. The point chamber is inserted at z=0 in the phantom and the phantom is translated as the chamber integrates the exposure from one end of the phantom to the next[8]. The exposure captured is called $D_{eq}$. This $D_{eq}$ is itself equivalent to the peak of the MSAD as described by Shope et.al in their original paper [9], and the use of a large number of scans by Shope et al [9]can be interpreted to imply an infinitely long phantom. It is also similar to the dose descriptor $D_{max}$ as defined by Spokas[4]. Another approach involves using a longer pencil chamber [10, 11] that can collect signals over the wider scatter tails. The longer pencil chamber is inserted into a phantom of 45 cm length. Due to the inconvenience of using 45cm long phantoms that are quite heavy, and the lack of longer pencil chambers, this approach has been slow to catch on. Since 2005 there has been a lot of effort put into advancing the use of the point chamber as the standard for defining the dose in CT due to the approach to equilibrium idea [12-14]. Beside the attention that the point chamber has received in these papers, there has also been a renewed interest in development of a mathematical description of the dose profile in CT. Most of the current mathematical descriptions of the dose profile in CT follow the approach of Gagne [3] of combining a scatter function with a primary beam function. Scatter functions have taken the form of a Lorentzian [3, 15], a sum of a two Gaussian functions [9] or a bi-exponential function [10, 14, 16] or other complicated functions [16]. The scatter function is combined with a primary beam profile function to get the complete dose profile function for a CT scan. These primary and scatter functions are typically combined in various ways or used by themselves to describe the total dose profile of a CT scan. For example, a scattered double Gaussian was used by Shope et.al in their original paper [9]. A Gaussian or Lorentzian function was used by Gagne [3] and also by Tsai et.al [15], and a sum of the primary beam and the scatter (a Gaussian or exponential) was used by Dixon and Boone [16] [12]. These complete dose profile functions are used to match the shape and width of the dose profile distributions. These dose profile distributions can be used to obtain the computed tomographic dose index CTDI if the profile can be shown to fit actual dose profile data from a CT scanner. With these fits in hand, a theoretical analysis of the dose to a phantom can be carried out. These theoretical analysis typically involve the use of a convolution approach [3, 10, 14, 16]. In addition to the convolution approaches, Monte Carlo formulations [12, 16] have often been used as further validation of the theoretical models. The convolution approach follows the approach described by Gagne earlier [3]. A common feature of the convolution approach and other approaches is to combine the primary function with an unknown function or some other scatter function. In the convolution approach, the scatter acts as a smoothing kernel. Other approaches to obtaining the scatter function involve derivation from LSF of the tube anode [14]. A more detailed scatter function was used in [16]. All these functions together with bi-exponential, Gaussian, Lorentzian and double Gaussian functions all can lead to fits of the dose profile in CT phantoms quite accurately, especially in the middle regions. In the tail regions, these functions fall off too quickly to describe the scatter tails. Despite the success of these functions, there is not one that holds dominance over the others. There are other possible candidates that can, either by themselves be used as scatter profiles or, be combined with a primary profile function to describe the complete dose profile in CT. In this paper, a model for describing the complete dose profile in CT is developed and presented. Although the motivations for the previous papers and this paper are similar, the functions, mathematical approach, and formulations are different. In particular, this paper uses several solutions to Riccati's equation to model the dose profile in CT. In addition, the primary beam is modeled as a trapezoidal function. Furthermore, we model and analyze combinations of the trapezoidal function, constructed functions, solutions to the Riccati equation and Gaussian functions as viable options to describe the complete dose profile measured by a detector when the CT beam is scattered in a phantom.

In this paper we use the product of the primary and scatter to represent the total dose profile in CT for all fits and simulations in the collimated region. Outside the collimated regions the fit involves summing the product of the primary and scatter functions with the scatter function. This article is divided into two parts. In the first part, we first develop and present a detailed mathematical model that describes the scatter dose profile in CT by using the solutions of Riccati's differential equation. We then model the primary beam by using two approaches both of which give viable approximations to the primary beam. First, we use the trapezoidal function to model the primary beam together with its penumbra region. Second we construct a function based on the use of the edges of the trapezoidal function. The top of the constructed function uses the analytic approximation to the Heaviside function to give it some curvature. Both approaches produce primary beams that are combined with the scatter beam to model the dose profiles presented below. The model is validated, especially in the collimated region, by fitting our complete model equation, containing both the scatter and primary components, to data from the o-arm cone beam system (O-arm Imaging). Outside the collimated region the scatter tails are modeled since the actual o-arm data did not include enough scatter data in that area.

In the second part, we present experimental data and simulations that connect our theory, developed in part 1 of this paper, to the use of the CTDI concept for measuring dose in a Stationary Cone Beam CT (SCBCT) system like the o-arm inter-operative scanner. We also formulate mathematical descriptions that reconcile the use of either a farmer chamber or a pencil chamber for collecting exposure data in the usual CT phantoms. We extend our formulation involving the use of an integrating detector or chamber to conventional MDCT scanners where there is table movement. Finally, we present an alternative approach to the TG111 approach that allows the use of the pencil chamber for MDCT. This approach is compared to experimental data and is found to be accurate to within 5% of data collected using the 'point' or farmer chamber approach in TG111.

# PART 1: DEVELOPMENT OF THE MATHEMATICAL MODEL

## A. MODELS OF THE SCATTER PROFILE IN CT

### A1. Riccati's equation as a model for the scatter dose profile in CT

In recent years there have been various functions that appear to represent the scatter dose profile in CT. In the original paper by Shope et.al [2] , two Gaussian functions were used to fit the dose profile, Others have used the bi-exponential function [10, 12]as well as a purely Gaussian function[5] . Due to the multiple possibilities of functions that are potential fits for the dose profile in CT, one can conclude that this problem is not a simple problem of finding a single integral in a finite dimensional vector space. In this paper, we introduce Riccati's equation as a model for the scatter dose profile in CT. We develop the theory in a Banach space with regards to the series approximation of the Riccati equation solutions  so as to lay a solid mathematical foundation for our theory of CT scatter dose profiles. The development of our solution in a function space is to allow us to single out from a group of functions the solution that best fits the data from the o-arm imaging system and can be applied to other MDCT systems. To this end, we start with the exploration of the solutions of the scalar Riccati type equations as a model for the scatter profile in CT phantoms.

Riccati's equation is a nonlinear ordinary differential equation (ODE) that has applications in various fields [17-21]. One type of scalar Riccati equation can be written in general form as

$$y' = h(z)y + g(z)y^n \ \ n \geq 1 \tag{1}$$

For n=1 this equation is soluble as a linear differential equation [22]. Generally, Riccati's equation can be transformed into a second order linear ODE. For our purposes, we modeled the scatter contribution to the dose profile in CT for the case where

$$h(z) = 0, \quad g(z) = z^m, and \ n = 2$$

This results in the following Riccati equation

$$y' + z^m y^2 = 0 \ for \ m \ \geq 1 \tag{2}$$

We first investigate the solution $y(z)$ of the above problem. $y(z)$ is an unknown output corresponding to an input z with domain belonging to the set of real numbers in $R^1$. The co-domain $\Omega$ is a subset of the Euclidean space $R^1$. The graph of $y(z)$ is represented by

$$\Omega \times \Omega = \left\{ (z, y(z)) : y \in R \times R, y' + z^m y^2 = 0 \ for \ m \ \geq 1 \right\}$$

Suppose that $y(z) \in C^1[a,b]$. Let $a, b \in R$ and $a < b$. We consider this to be the space comprised of those functions $y(z): [a,b] \to R$ that are continuous on the open interval $[a,b]$, and have a single derivative in z. The function $y(z)$ represents the solution to the scalar Riccati equation, or some suitable equation, and will be used to describe the total scatter profile in CT as measured by a detector.

If we consider the detector to be a pencil integrating chamber, then we can say we are interested in the total amount of charge collected by the pencil chamber. The magnitude of charge is described by the area of the region under the graph of a function $f_T(z)$ consisting of the scatter component $y(z)$. We start off by considering the function $y(z)$ to be the solution of equation (2) above and let $y(z)$ be defined in some co-domain $\Omega \in R^1$. Then

$$y(z): \Omega \times \Omega \to R$$

Is a mapping. We say that $y(z) \in L_p^1(\Omega)$ if $y(z)$ has derivatives of order up to 1 in z and $1 \leq p < \infty$. Its metric (norm) is defined by

$$\|y\|_p = \left( \int_\Omega \|y\|^p \, d\mu \right)^{\frac{1}{p}}, \quad y \in C^1[a,b]$$

This is a metric or norm defined on the function space $C^1[a,b]$. If we choose p=2, then this makes $L_2^1(\Omega)$ a Hilbert space of square integrable functions. In this paper we will work in the Banach space where $p \neq 2$, so as to keep it general. Note that all Hilbert spaces are also Banach spaces. However, the inverse is not true.

Besides the class of equation in (2), we can also define a second class of Riccati equations where

$$h(z) = 0, \quad g(z) = z^m + z^{m-1}, n = 2, \quad m \geq 1$$

$$y' + (z^m + z^{m-1})y^2 = 0 \; for \, m \geq 1 \quad (3)$$

## A2. Analytic solutions of Riccati's equation

Analytic solutions of the above equation (2) are condensed into the following,

$$y(z) = \frac{(m+1)}{((m+1)C + z^{m+1})} \quad (4)$$

Where $m$, is an integer. The choice of $m$ leads to a spreading of the profile of the function $y(z)$. The solution has a general form that can be applied as a model to the scatter dose profile in CT scanning with a chamber placed at the z=0 position, i.e. $y(0)$. Traditional pencil chamber was designed for a single axial scan at z=0 where the entire dose profile will be with the chamber endpoints. In (4), the constant term $(m+1)C = B$. The domain is the set of all real numbers $\{z \in R : -B < z < B\}$, with a co-domain $(y \in R : 0 < y \leq 1)$--- we have normalized the output $y(z)$.

For m = 1 we can rewrite equation (4) above using B as defined above

$$y(z) = \frac{2}{(B + z^2)} \quad (5)$$

Where $B = (m+1)C = 2C$ for $m = 1$

Based on this result, we developed an improved analytic model that led to a more exact version of our Riccati equation. This was used for fitting the data. This is shown below

$$cy' + azy^2 = 0$$

The solution to this equation is

$$y_s(z) = \frac{2a}{(b + cz^2)} \quad (6)$$

The parameters a, b, and c were adjusted to fit the real data. Note that $b = aB$ , where $b$ is related to the parameters that spread or narrow the scatter profile, such as collimation, scan length and phantom dimensions, and c is related to parameters that reduce the magnitude of the profile such as attenuation. A nonlinear regression analysis was done using Solver in Excel (Microsoft Corp. Bellevue Washington) to improve the fit. The parameters used for the fit are reported in the results section for both the head and body phantom data for the O-arm cone beam system.

A second class of Riccati's equation (3) is shown below

$$y' + (z^m + z^{m-1})y^2 = 0 \ for \ m \geq 1$$

This second class of Riccati equations can also be solved analytically. For this class of Riccati equations, there is a solution of the form

$$y(z) = \frac{m^2 + m}{((m^2 + m)C + (mz + m + 1) z^m} \ for \ m \geq 1 \qquad (7)$$

Where $m$, is an integer. The choice of $m$ leads to a spreading of the profile of the function $y(z)$. We considered the solution for $m$ up to 6 to demonstrate the generality of the solution. It is certainly possible to go to higher values of $m$ if the situation dictates it. These solutions all exist in the function space defined above and can be used as approximations to the scatter dose profile in CT. For our purposes we adopted, for simplicity, the first class of Riccati equations to use for modeling the scatter profile in CT.

## A3. Series representation of $y(z)$

The basic solution obtained (5) above can be written as a series representation as below

$$\frac{1}{1 - (-\frac{z^2}{B})} = 1 - \frac{z^2}{B} + \left(\frac{z^2}{B}\right)^2 - \left(\frac{z^2}{B}\right)^3 \ldots$$

$$y(z) = \sum_{n=0}^{\infty} (-1)^n \left(\frac{z^{2n}}{B^n}\right) \qquad (8)$$

## A4. Convergence of the series representation

This series can be shown to converge for the following two cases. For z approaching infinity, the function approaches zero, while for z approaching zero, the function become a finite number. Both of these two cases is shown below

## A5. Series expansion at $z = 0$

Series expansion of the function $y(z) = \sum_{n=0}^{\infty} (-1)^n \left(\frac{z^{2n}}{B^n}\right)$ is the function

$1 - \frac{z^2}{B} + \left(\frac{z^2}{B}\right)^2 - \left(\frac{z^2}{B}\right)^3 \ldots$. This converges to 1 as we expect for the limit as $z \to 0$. This just tells us that at the location z = 0 where a probe might be placed, the maximum relative signal should be collected. This convergent series can be considered to be a Cauchy sequence in the space described.

# A6. Series expansion at $z = \infty$

The series expansion at $z = \infty$ of the function y(z) is given as $\left(\frac{1}{z}\right)^2 - \frac{B}{z^4} + \frac{B^2}{z^6} + O\left(\left(\frac{1}{z}\right)^7\right)$. This function converges to zero as we expect. This tells us that a probe will not read a signal if the scan dose profile was placed an infinite distance from the probe. The convergence of these functions in the limits at zero and infinity make the space complete. We earlier defined a norm for the space $L_p^1(\Omega)$. A complete normed space is a Banach space. It must be noted that Lorentzian, Gaussian and exponential functions also show the same convergence properties at $z \to 0$. An exponential function can be written in compact form as $e^{-x} = \sum_{k=0}^{\infty} \frac{(-x)^k}{k!}$. A Gaussian is written as $e^{-x^2} = \sum_{k=0}^{\infty} \frac{(-x^2)^k}{k!}$. The Lorentzian form $\frac{1}{1+x^2} = \sum_{k=0}^{\infty}(-x)^{2k}$ is the same, except for a constant term, as the form obtained for the solution of Riccati's equation. These functions all satisfy the same convergence properties as the solution $y(z)$, obtained from the Ricatti equation. Gaussian and Lorentzian functions are commonly used in spectroscopy. They can be combined by convolution to give a Voight function. In spectroscopic analysis, a Voight function, which is a convolution of the Gaussian and Lorentzian functions, can be used to extract the relative contributions of these two functions to a spectral line shape function[23, 24].

## Definition 1. Scatter function in a medium

We now define the scatter function $y(z)$ as the solution to the non-linear ODE called the Riccati equation; $y' + z^m y^2 = 0$. The solution converges to zero at infinity and is maximum at z=0. The convergence properties are shown in sections 1.6A and 1.7A above.

## Theorem 1.1 (scatter beam profile)

If $y(z)$ is a unique well posed solution in some domain $\Omega \in R^1$, and the graph of $y(z) = \{(z, y(z)) \in R \times R : -B < z < B, y' + z^m y^2 = 0\}$ $for$ $m = 1$, and the solution is $C^1[-\infty, \infty]$ on the open set containing R. Then

$$y(z) = \frac{2}{(B + z^2)}$$

Is the solution to a non-linear ODE called the Riccati equation. It is also called a scatter function.

## Corollary: Convergence, maximum and minimum norm

The maximum value of the function is $y(0) = sup_{z=0}\|y(z)\| = 1$. And the minimum value is described as; $y(\infty) = min_{z=\infty}\|y(z)\| = 0$

The proof of theorem 1.2 is obtained by solving the nonlinear ODE equation 2 for m =1. The solution of the Riccati equation

$$y' + z^m y^2 = 0 \ for \ m \geq 1$$

Is solved analytically first by setting $\frac{y'}{y^2} = -z$ . The solution follows from this by using basic calculus to get

$$y(z) = \frac{2}{(B + z^2)}$$

Where $B = (m + 1)C = 2C$ for $m = 1$. Note that this solution is identical to the Lorentzian line shape profile although it was obtained from solving a nonlinear ODE, the Riccati equation.

# B. MODELS OF THE PRIMARY BEAM PROFILE IN CT

## B1. Trapezoidal function as a model for the primary beam profile in CT

In previous work, the primary beam was described as a unit rectangular function ([3, 10]). In this paper, we consider the ideal primary beam profile in CT to be a trapezoidal function. It is useful to use the trapezoidal function in this case since sloping regions of the trapezoid legs can fall outside the slices used for acquisition (A). The reconstruction is done in the region (nT). The trapezoid function can also be modeled to have a penumbra region and is shown below in Fig 1. We can also include the heel effect in the model. Our use of the trapezoid is an idealization of a primary beam in air. We consider the scatter to be a modulation of the primary beam once in the phantom.

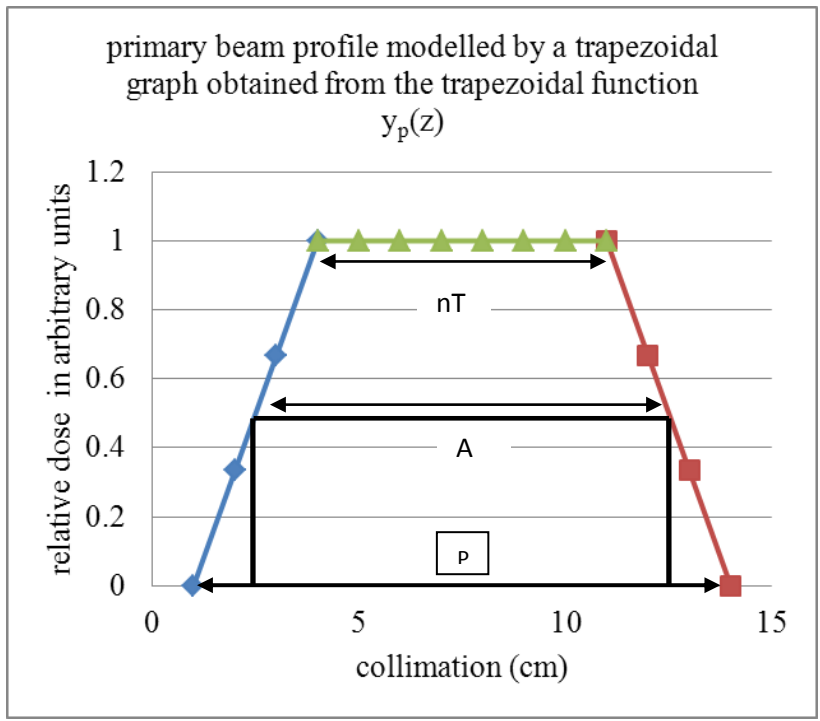

Fig1. A trapezoidal function used to simulate the primary beam in CT scanning. The region used for acquisition is A. The reconstruction region is nT. The FWHM or aperture will fall in the region called A. Where p is the total dose distribution.

In this simulated graph of a primary beam profile,

$$y_p(z) = KF_0 g \begin{cases} \frac{z-a}{b-a} & 1 \leq z \leq 4 \ (penumbra) \\ 1 & 4 \leq z \leq 11 \ (collimated\ region\ ) \\ \frac{d-z}{d-c} & 11 \leq z \leq 14 \ (penumbra) \end{cases}$$ where the collimation is $b \leq z \leq c$,

and the penumbra region is given by the defined edges as above. For a collimation of 7mm one would set the collimation range described above together with the penumbra regions defined. The figure described above is not to scale. The first term outside the curly brackets is a term containing a constant K, the exposure $F_o$, geometric efficiencies g. The above primary beam model can be used together with the scatter profile obtained from solutions to the Riccati equations to model the total dose profile $f_T(z)$ in CT. These results are presented below. Besides the use of the trapezoidal function, we also discuss the possibility of other models of the primary beam in the section below.

## B2. Constructed function as a model for the primary beam profile in CT

We also model the primary beam by using a modified version of the analytic approximation to the Heaviside step function as well as constructed functions defined in specified regions Fig 2. The width of the primary beam is determined by the collimation we set. But since the beam itself has a penumbra region, the FWHM is different from the actual collimation displayed. This accounts for the different efficiencies of the various collimations. We define the function as follows.

$$y_p(z) = \begin{cases} mz + M \ if \ z < D(penumbra) \\ \left(\frac{H}{(1+e^{(z-D)})} + \frac{H}{(1+e^{(D-z)})}\right) if -D \leq z \leq D \ for \ D \in R \ and \ D = -8cm \ and \ D = 8cm \ (collimated\ region) \\ -mz + N \ if \ z > D(penumbra) \end{cases}$$

Where ($H = KF_0 gL$). This function, which is a mixture of the analytic continuation to the Heaviside function and the constructed functions, is used to generate the primary beam profiles used for both the head and the body phantoms shown in the graphs 9 & 10 below. For dosimetry purposes, the integral z is over the length of the detector, or the scan length depending on whether the detector is long enough to cover the scatter tails or whether a "point" detector will be used. In the case of a point detector, the scan length should be equal to the length of an integrating chamber [25] if the two signals are to be compared. More will be said on this later.

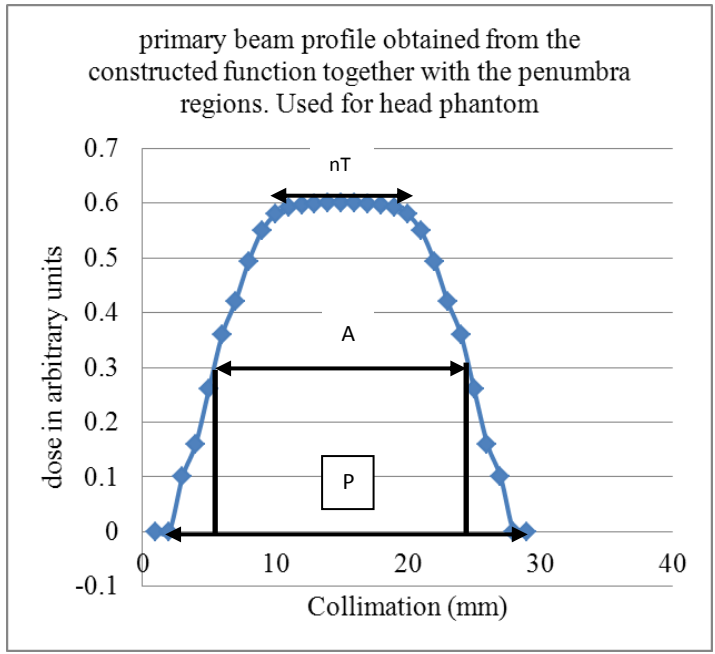

Fig 2. A simulated primary beam profile based on the analytic approximation to the Heaviside step function and a constructed function. The acquisition region is A, the reconstruction region is nT, and the total profile is P.

## Definition 2.

The primary beam profile is a trapezoidal function with penumbra edges, and is defined by the collimator and the bowtie shaping filters. The intensity is determined by the energy of the beam, the number of photons and the filtration in the beam. The trapezoid becomes more triangular as the collimation gets more narrow.

# C. MODEL OF THE COMPLETE DOSE PROFILE IN CT FOR A BEAM ATTENUATED IN A PHANTOM

## C1. Combinations of the primary and scatter models

The complete dose profile in CT can now be modeled accurately at least in the collimated region by taking the product of the modeled primary beam and scatter profiles obtained from the Riccati equation. The complete profile including the collimated region and the scatter tails are modeled by adding the product of the primary and scatter component to a scatter component, equation (9) below. Several graphs of the fits to O-arm data are shown in the results section below. The O-Arm has been previously described by Zhang et.al [26]. A nonlinear regression analysis was used to improve the initial fits to the o-arm data. The analysis was done using the Solver feature in Microsoft Excel. The fit parameters obtained are in table 1 below. The complete dose profile distribution, with primary and scatter components is now shown as below;

$$f_T(z) = y_p(z)y_s(z) + y_s(z) = f(z) \tag{9}$$

Where $y_p(z)$ is given by the trapezoid primary function or some other suitable primary function; we also consider a constructed primary function; and $y_s(z)$ is given by the solution to Riccati's equation, $y_s(z) = \frac{2a}{(b+cz^2)}$. The scatter function may also be considered to be a Gaussian distribution of the form. $y_s(z) = e^{-\frac{(z-m)^2}{zk^2}}$. Or it may be obtained from a blurring of the primary function with the bi-exponential or a Lorentzian line shape function.

## C2. The modeled dose profile distribution in SCBCT of the O-arm system

The dose profile in CT can now be modeled accurately for the collimated region by taking the product of the primary and scatter profiles obtained from both the Riccati equation and the trapezoidal or constructed function. By adding the scatter component to the produc of the primary and scatter, the scatter tails outside the collimated region can be modeled. The fits are improved by using a non-linear regression analysis in excel (Microsoft, Bellevue, WA). This approach to modeling the beam yielded excellent fits to the o-arm data as shown in figures 9 - 13 in graphs for the collimated region as well as the scatter tails. From the graphs it can be seen that for this wide cone beam (15.7cm at iso-center) a 450mm phantom is not enough to cause the scatter tails to converge. A larger phantom is necessary as was shown by Mori et.al [11]. The scatter tails in the graphs 9 -13 are modeled by adding a scatter component to the product of the primary and scatter components. The o-arm data represents the exposure profile $f_T(z)$ collected by a farmer chamber. Based on the results obtained from fitting the graphs of the o-arm data to the model, we can define the total beam profile as below.

### Definition 3.

The total beam profile in a phantom is the product of the primary beam and the scattered component plus the scattered component as shown below.

The total CT beam $f_T(z)$ may be described by the function below

$$f_T(z) = \begin{cases} y_p(z) & \text{primary beam (no scatterer)} \\ y_p(z)y_s(z) + y_s(z) & \text{beam in phantom} \end{cases} \tag{10}$$

## C3. Connection to Dosimetry for MDCT

In order to perform dosimetry we must define a few conditions about the dose profile and the phantom to be used. For the exposure to a patient to be accurately reflected by the exposure to a detector placed in a phantom, the detector must collect all signals generated in the phantom. Thus the conditions for dosimetry to be performed are that the phantom must be infinitely long, and the dose profile must converge to zero at the ends of the phantom. These two conditions can be more clearly defined below

## Definition4. (Scatterer and its dimensions)

The Scatterer is a phantom long enough to allow the scatter tails to converge to zero at distal ends of the phantom. This requirement for convergence is necessary and sufficient to conclude that the phantom size is large enough for the collimation used. A check for this can be done by taking a measurement at one end of the phantom with a 'live' detector. If a reading of close to zero is obtained, we can say that the scatter tail is zero at that point. This of course depends on issues with sensitivity and response of detector e.t.c.

## Definition5. (Equilibrium or peak dose $D_{peak}$(z=0).

Having an infinitely long phantom allows us to measure the peak dose at a single location z=0. This is because the contribution of signals from distances far away from the detector placed at z=0 will be close to zero, so the dose measured at z=0 will be the peak dose. The peak dose is also called the equilibrium dose $D_{peak}$(z=0) ([10]). Based on this definition, a useful synthesis of the relationship between the dose profile and the phantom length is given below as a necessary and sufficient condition.

## Definition6. (Necessary and Sufficient condition)

If the total beam profile $f_T(z)$ converges to zero at infinity inside a Scatterer, then we can say that the scatterer is infinitely long. The convergence to zero at infinity of a dose profile $f_T(z)$ in a scatterer is both a necessary and a sufficient condition to conclude that the scatterer is infinitely long or large. Note that this statement is also true for the case where the infinite length of the scatterer is both the necessary and sufficient condition for the dose profile $f_T(z)$ to converge to zero at infinity.

## Theorem 1.2 (convergence of the scatter profile)

We can then state the theorem as this. The dose profile $f_T(z)$ converges to zero at infinity if and only if the scatterer is infinitely long. Or the scatterer is infinitely long if and only if the dose profile $f_T(z)$ converges to zero at infinity. Note that a long or infinite scan length, used by Gagne [3]by definition also falls under the concept of an infinitely long scatterer. A corollary to the theorem about convergence and an infinite scatterer is given below

## Corollary (Equilibrium or peak dose $D_{peak}$(z=0):

 If the dose profile $f_T(z)$ converges to zero at infinity, then we can say that the dose D(z) measured by a detector placed at the z=0 location in a phantom is the equilibrium dose $D_{eq}$(z=0). This can more accurately be called peak dose at z=0.

## Proof of Corollary:

We can prove this as follows. Suppose a detector is placed at z=0 and is required to measure a signal from a function that tails off. If the function is directly over the detector, it measures the maximum signal. If the function is placed at a distance equal to half the FWHM from the detector it measures half the maximum signal. At another ½ FWHM, the signal is now approaching zero. The farther the function is from the detector the smaller the signal the detector will measure. It is clear that the signal measured by the detector at its location will get closer and closer to zero as the function is moved farther away. The signal will reach a peak value and level off. i.e. and equilibrium is reached between the peak signal and distance of the function from the detector. This completes the proof.

We are now ready apply these ideas to a formulation of CT dosimetry that can be used to understand the relationship between the exposure measured, and detector size.

It must be pointed out that once a detector is inserted into the beam to measure exposure, the exposure collected by the detector is the area of the region under the graph of a segment $f_i(z)$ of the dose profile that falls on the detector. Note that $f_i(z) \equiv f(z_i)$ in the following analysis. The total exposure is the sum of each area segment over the whole dose profile from one scatter tail to the other. The total exposure is therefore the integral of the total dose profile function $f_T(z)$ which can then be used to calculate a dose or dose index according to the detector length.

If the detector is long enough to cover the entire dose profile including the scatter tails, and the phantom is long enough to allow the profile scatter tails to converge to zero at the ends of the phantom, then a single integral for $f_T(z)$ can be used. If on the other hand the detector is not long enough to cover the entire scatter tails, then the detector must be used to integrate the entire function $f_T(z)$ by summing the areas of the region under the graph.

Now, assuming we have met the conditions defined above in theorem 1.2, we can now develop a theory around the use of a detector to measure exposure once the detector is inserted into the phantom. Let's consider For example a detector that covers only 1/n of the total dose profile function $f_T(z)$. This detector can be summed n times as shown below by moving the detector to n locations in a suitable phantom (theorem 1.2) i.e. from one end of the dose profile to the other for each axial scan. In general, we consider a two dimensional region of the graph bounded by $f_i(z)$ and the interval $\Delta z_i$ ( $z_i$ to $z_{i+1}$) as shown below. The exposure ($\Delta D_i(z)$) collected by a detector placed at that point of the graph is the area of the bounded region. Geometrically, the integral is the area of the region under $f_i(z)$ and $\Delta z_i$ is represented as shown below;

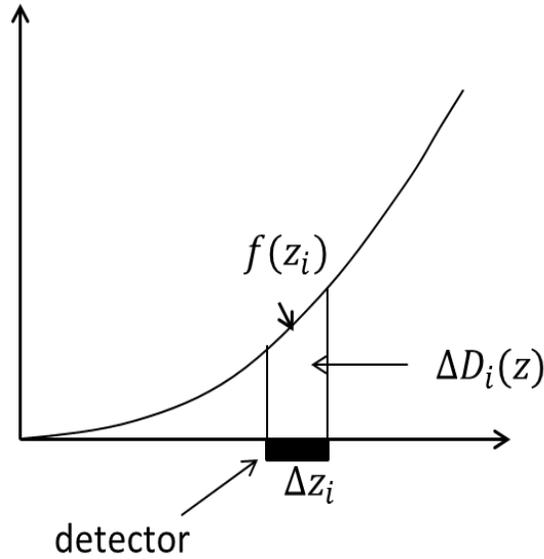

$$\Delta D_i(z)_{area} = f_i(z)\Delta z_{i\,detector\,length} \qquad (11)$$

$$dD_i(z) = \lim_{\Delta z_i \to 0} f_i(z)\Delta z_i = \int_{-l/2}^{l/2} f_i(z)dz$$

$$D_i(z) = \int_{-i}^{i} \Delta D_i(z)$$

the endpoints are the dimensions of the detector. We can then write

$$D_i(z) = \int_{-l/2}^{l/2} f_i(z)dz$$

$$D_{total}(z) = \sum_{i=1}^{n} D_i(z)$$

For a detector that is 1/n times the length of the dose profile, we can sum n integrals as shown

$$D_{total}(z) = \sum_{i=1}^{3} \int_{-l/2}^{l/2} f_i(z)dz_{i\,detector\,active\,lenght}$$

$$D_{Total}(z) = D_1(z) + D_2(z) + \cdots + D_n(z)$$

In terms of an integral, we have

$$D_{Total}(z) = \int_{-l/2}^{l/2} f_{T1}(z)\,dz_1 + \int_{-l/2}^{l/2} f_{T2}(z)\,dz_2 + \cdots + \int_{-l/2}^{l/2} f_{Tn}(z)\,dz_n$$

This can be done for different combinations of our primary beams and the scatter beams mentioned. This approach was used to fit the graphs of the various primary and scatter beams. Where the integral over z is over the length (l) of the detector or chamber used for measuring the signal(exposure). The length of the detector and the length of the phantom determine the number of integrals needed to collect the entire exposure from one scatter tail to the other. The equivalent situation of placing a detector in one location (z=0) and translating the phantom through the scan [27]is described later.

# C4. Extension of the theory to SCBCT

We can extend the above theory to a stationary cone beam CT (SCBCT) system. Recall that the o-arm is a true SCBCT system, in that there is no movement of the table during scanning. The beam is 15.7 cm at the iso-center of the bore. The theory developed in this article, while it was modeled to the data from an o-arm SCBCT system, can also be applied to the MDCT system with wide beams that approach the width of the o-arm system. The Toshiba 320 slice Aquilion One system is one such system. For a single axial scan at the z=0 location in a long phantom, a pencil chamber (100mm) can integrate the complete dose profile and give the dose to the region irradiated according to the formula below

$$D_{pencil}(z) = \sum_{i=1}^{N} f_i(z)\Delta z_{ipencil}$$

$$CTDI_{100N} = \frac{D_{pencil}(z)}{nT}$$

This formula allows us to fix the phantom in one place and move the chamber from one location to another through the phantom for each axial scan until the entire exposure profile is integrated. This approach was used to collect the o-arm data [26]and also to collect dose profile data for the Toshiba Aquilion system[28] although the mathematical analysis described here was not shown in either case.  It must also be pointed out that in the article by Lin et.al [28], the 2mm solid state detector was moved 10mm/s each second resulting in a pitch of 5 for each measurement. In the theory developed in this paper, the number of locations of the chamber is determined by dividing the phantom length that allows the scatter tails to converge to zero at the ends according to the necessary and sufficient condition, by the detector active length. This allows us to use the usual pencil chamber detector or farmer chamber detector with a large phantom. Obviously, this can be done for other MDCT scanners with smaller beam collimations. Others have argued that the using the FWHM is more appropriate that the collimation nT in the CTDI formula. Oliveira et.al [6] defined $CTDI^*$ where nT is replaced by the FWHM, while Dixon and Ballard [8] defined the $CTDI_a$, where 'a' refers to aperture. Aperture is equivalent to the FWHM.

Exposure readings are taken at the peripheral and central locations and a weighted CTDI is obtained. The volume CTDI can then be calculated the usual way.

## Definition 6. Total Beam profile in CT scanning

We will now like to define the total beam in CT as follows. The total CT beam D(z) is the exposure detected by summing the signals from an integrating detector placed at multiple locations along the entire length of the beam profile. The number of locations depends on the length of the detector used for measuring the signal and the distribution of the beam profile. This applies to a profile in air or in a scatterer.

## Detector length and signal collection

The total signal collected by an integrating detector D(z) depends on the length L of the integrator in a linear fashion. Longer detectors will collect more signal than shorter detectors. Efficiency is given by ε. Efficiency is better for wider collimations that smaller ones. In air we can define the total signal as below.

$$D_{air}(z) = \varepsilon \alpha k f_T(z) g l$$

where $l$ is detector length, and $f_T(z)$ has the meanings in equation (10). In a phantom (see equation 10), the signal collected is proportional to the area under total dose profile $f_T(z)$ that the detector can collect. This can be shown as;

$$D(z) = \varepsilon \alpha k g \int_{\frac{-l}{2}}^{\frac{l}{2}} f_T(z)\, dz.$$

## Theorem1.3. Summing theorem

Let $f_T(z)$ be a function that describes the total beam profile and fits the data. Then the area of the region under the graph of $f_T(z)$ is the exposure collected by a detector placed in that region under the graph. Summing segments of $f_T(z)$ over multiple locations over the entire distribution of the profile gives the total exposure $D(z)$. This can be represented algebraically for a pencil shaped chamber and a farmer chamber as

$$\text{D}(z) = \sum_{i=1}^{N=3} f(z_{ipencil}) \Delta z_{ipencil} = \sum_{i=1}^{N=13} f(z_{ifarmer}) \Delta z_{ifarmer}$$

## Corollary (detector length $l$)

The length $l$ of the detector determines how much of $f_T(z)$ is integrated at each location. A longer detector will integrate a larger section than a shorter detector.

## Proof 1.3

The proof to this theorem is presented as data collected in table 1 below. The results from the table prove the statements of the theorem.

# C5. Scan length and detector length

An equivalent scenario to the one described above where a detector is moved from one point in a phantom to another in equal increments that allow for a complete integration of the dose profile during axial scans is described below. In this case, a detector is placed at the center of a phantom and the phantom is translated in equal increments from one location to another during scans so as to allow for a complete integration of the dose profile by the detector. The can be done for an axial scan or for a helical scan. Let $\Delta d$ be detector length and let $\Delta z$ be table movement per scan or scan length. Note that the over-ranging effect of the dose profile is accounted for since the detector captures the dose as it is distributed.

With the detector $\Delta d$ placed in the center of a suitably large phantom according to theorem 1.2, each scan results in the detector integrating a segment of the dose profile as shown in figure below. Table movement per scan is given by

$$\Delta z \ (mm) = table \ speed \ \left(\frac{mm}{rot}\right) \div time \ per \ rotation \ \left(\frac{s}{rot}\right)$$

For axial scans we can write

$$\Delta z \ (mm) = collimation \ \left(\frac{mm}{rot}\right) \div time \ per \ rotation \ \left(\frac{s}{rot}\right)$$

For axial scans, multiple contiguous scans will be used in order to translate the phantom from one edge of the dose profile to the other.

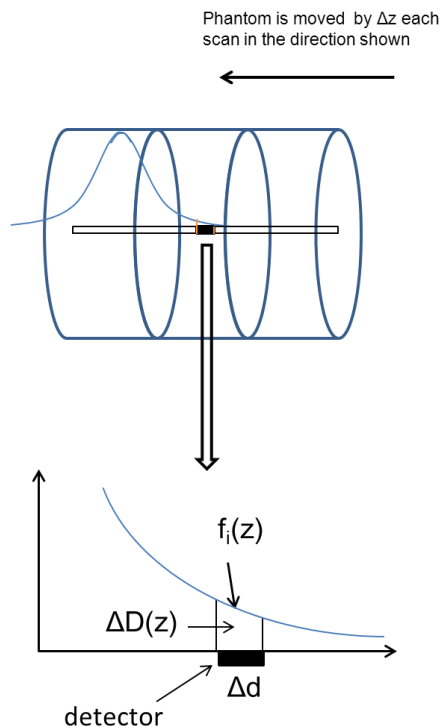

Phantom is moved by Δz each scan in the direction shown

The dose profile for each segment that falls on the detector is given by $\Delta D_{id}(z) = f_i(z)\Delta d$. This is an area $A_d$ that represents the exposure $\Delta D_i(z)$ collected by the detector at that location. The area of the region under the graph of $f_i(z)$ that is due to the phantom translation $\Delta z$ is given by $A_p = f_i(z)\Delta z = \Delta D_{ip}(z)$. Where we note that the $f_i(z)$ is the same segment of the dose profile graph that is in the location of the detector. For the number of photons collected by the detector in a region to be equal to the number of photons actually in that region, the area of the two regions should be equal,

$$A_d = A_p$$

$$f_i(z)\Delta d = f_i(z)\Delta z$$

$$\Delta D_{id}(z) = \Delta D_{ip}(z)$$

This result is the same as the result obtained above (11) where the phantom was kept fixed in the beam and the detector was moved from one end of the dose profile to the other. The rest of the analysis follows the same way. An interesting aspect of this result is that the detector length $\Delta d$ should be equal to the table movement per scan, $\Delta z$, for each translation during axial scans. This will allow the detector to integrate the entire dose profile from one end to the other in a contiguous fashion. The exposure for a region under the graph of the function $f_i(z)$ can be written as $D_i(z)$. The total exposure from integrating from one end of the dose profile to the other is again given as

$D_{total}(z) = \sum_{i=1}^{n} D_i(z)$ or in terms of the integral is can be written as

$$D_{total}(z) = \sum_{i=1}^{n} \int_{-l/2}^{l/2} f_i(z) dz \qquad (12)$$

Let's now consider the following two cases during an axial scan. The first case is where the detector dimension is smaller than the table translation per scan. The second case is where the detector dimension is larger than the table translation per scan. The first case is below;

1. $\Delta d < \Delta z$

If the table movement per scan, $\Delta z$, is larger than the detector length, we can expect that the detector will underestimate the total exposure over the entire dose profile $f_T(z)$ due to undercollection of the signal in the region. This can be seen from the analysis below

$A_d = f_i(z)\Delta d$ is the exposure the detector receives for a segment of the dose profile function.

$A_p = f_j(z)\Delta z$ is the actual area of the region created by the movement of the phantom and represents the number of photons in that area.

Since $\Delta z$ is larger, then the segment of the dose profile will be larger. Instead of $f_i(z)$ we write $f_j(z)$ without knowing the relationship between the larger segment $f_j(z)$, and the smaller segment $f_i(z)$. We can also say $\Delta z = n\Delta d$. Using this formula and the area formulae above, we arrive at the result below

$$A_p = n \frac{f_j(z)}{f_i(z)} A_d$$

We see that the detector underestimates the photons in the region bounded by the table movement $\Delta z$ and segment of the dose profile $f_j(z)$ by the factor

$$\frac{f_i(z)}{n f_j(z)}$$

In essence, if a we replaced our detector of length $\Delta d$ with one of length $\Delta z$ , then it will collect an exposure equal to $D_p(z) = f_j(z)\Delta z$. In terms of exposure notation we can write

$$D_d(z) = \frac{f_i(z)}{n f_j(z)} D_p(z)$$

2. $\Delta d > \Delta z$

If the table movement per scan, $\Delta z$, is smaller than the detector length, we can expect that the detector will overestimate the total exposure over the entire dose profile $f_T(z)$ due to the overcollection of each region during the scan. This analysis follows the analysis above in the same way.

Since $\Delta z$ is smaller, the segment of the dose profile will also be smaller. Instead of $f_i(z)$ we write $f_k(z)$ without knowing the relationship between the two other than that $f_i(z) > f_k(z)$. We can also have $\Delta d = n\Delta z$. Using this relationship and the area formulae we arrive at the result below

$$A_d = n \frac{f_i(z)}{f_k(z)} A_p$$

The detector oversamples the exposure by the factor $n \frac{f_i(z)}{f_k(z)}$ if the detector is larger than the table movement per scan during an axial scan. In exposure notation, we have

$$D_d(z) = n \frac{f_i(z)}{f_k(z)} D_p(z)$$

Both scenarios described above lead to a difficulty. The table movement in an axial scan is usually equal to the collimation set if the scans are to be contiguous. This means that it will be difficult to match the table movement to the detector length for smaller detectors like the farmer chamber. And even for a longer pencil chamber, if the detector is to be placed in the phantom and the phantom translated, then the detector length and table movement length should be matched as closely as possible. This can be resolved by using a helical scan instead of an axial scan. In an earlier article, Dixon proposed using a helical scan together with the farmer chamber placed at the z=0 position of the phantom as the phantom is translated through the beam. We analyze this scenario next

For helical scans, a single continuous scan will be used in order to translate the phantom from one edge of the dose profile to the other. It is useful to start by defining a pitch associated with the areas associated with the regions of the graphs falling on the detector and also the area due to the scan regions connected with the scan length.

$$pitch = \frac{Area\ under\ graph\ due\ to\ translation\ of\ phantom\ (\ A_p)}{Area\ seen\ by\ the\ detector\ due\ to\ the\ detector\ dimension\ \Delta z\ and\ the\ function\ f_i(z),(A_d)}$$

We can further make the extension, as we did in the cases above, that the area under the graph of the function is the exposure in that region. We arrive at the relation below for helical scanning

$$P = \frac{D_p}{D_d}$$

Or we can write it as below.

$$D_d = \frac{D_p}{P}$$

The same result can be arrived at by starting off as we did earlier and working through the algebra. We start by defining the exposure in the region under the graph of the function $f_T$ and the scan region covered by the scan length $l$

$$D_P(z) = \lim_{\Delta z_i \to 0} f_i(z) \Delta z_i = \int_{-l/2}^{l/2} f_i(z) dl$$

$D_p(z) = \int_{-l/2}^{l/2} f_T(z) dl$. We can also define the total dose profile as $f_T = \sum_{i=1}^{n} f_i(z)$. We further define the pitch P associated with the scan length and the detector dimensions as $= \frac{\Delta l}{\Delta z}$. With this in mind, we have $dl = Pdz$. From $D_p(z)$, above, we get

$$D_p(z) = P \times \left( \sum_{i=1}^{n} \int_{-l/2}^{l/2} f_i(z) dz \right)$$

This is the same formula as we obtained above for the total dose equation (12) except for the pitch P that multiplies the whole result. The formula in parenthesis is equal to the dose the detector measures i.e.

$$D_d = \left( \sum_{i=1}^{n} \int_{-l/2}^{l/2} f_i(z) dz \right)$$

$$D_d = \frac{D_p}{P}$$

If the pitch value is set as one, the results are the same as we would expect.

# PART 2: DEVELOPMENT OF THE EXPERIMENTAL MODEL

## A. PHANTOM DOSES IN CT MEASURED BY A DETECTOR

### 2A. Scan length and collimation

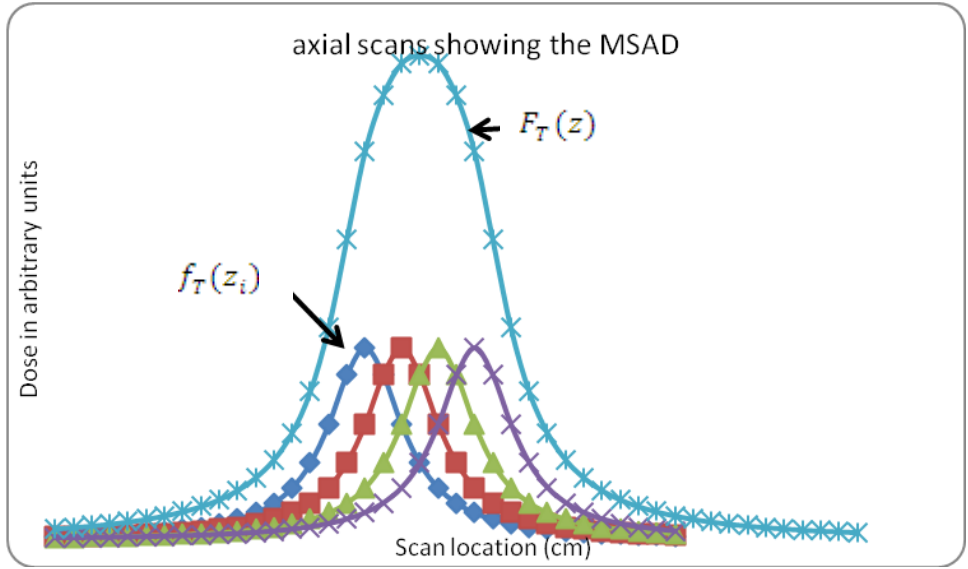

Fig 3.The simulated cumulated dose profile due to a series of simulated scans along a phantom. The total dose profile $F_T(z)$ is the sum of the individual dose profiles $f_T(z_i)$.

For the purposes of dosimetry we collect the exposure from a single axial scan for a given collimation. For axial scans, the scan length is equal to the collimation set. The table moves a distance equal to the collimation. For contiguous axial scans, the scan interval is equal to the collimation set, leading to the graph shown above for the MSAD. When a detector such as a pencil chamber is used, the goal is to collect the entire exposure from a single axial scan and to multiply this by the scan length to get the DLP. Since the exposure profile $f_T(z)$ in a phantom extends well beyond the collimation set, we need to either have a detector long enough to cover the entire profile, or contrive to measure the profile piece by piece.

In this section we consider the scan length L and its relation to the total dose profile $f_T(z)$ and also the connection to MSAD. This is described below.

For multiple axial scans, we can sum the function $f_T$ over the number of scans. Since each scan contributes scatter to the previous scan, the results is shown graphically above as the MSAD and is represented algebraically below.

$$D_T(z) = \sum_{i=n}^{N} D(z_i) = D(z_n) + D(z_{n+1}) + .. + D(z_{N-1}) + D(z_N)$$

$$MSAD = \sum_{i=0}^{N} CTDI_i = \sum_{i=0}^{N} \frac{D_T(z_i)}{nT}$$

Each $D(z_i)$ in this summation is the measured exposure from a single dose profile $f_T(z)$ by a detector that covers the entire dose profile including scatter tails, as described above in theorem 1.3. N is the number of scans.

## 3A. Phantom doses measured by farmer chamber using the TG111 approach compared to pencil chamber readings.

The experiment below demonstrates that the pencil chamber can be used to measure the dose from wider MDCT scanners such as the o-arm. The dose increases linearly with collimation and detector length and is only limited by the length of the detector; in this case a pencil chamber or a point chamber. Longer chambers will measure more instantaneous doses from wider MDCT scanners since they will integrate more of the dose profile each time. The simulation with a Siemens scanner supports this theory. For longer MDCT scanners, a pencil chamber can be used to measure the dose from several parts of the phantom and added to get the total dose. This dose can still be called CTDI$_n$ if the usual approach to getting CTDI is used i.e. central and peripheral readings to get the weighted CTDI and progressing the usual way. In this case, the subscript 'N' is the ratio of length of the entire convergent dose profile and detector length. If a 100mm active length pencil chamber is used and multiple readings are taken to integrate the entire dose profile $f_T(z)$, then

$$CTDI_{100N} = \sum_{i=1}^{N} CTDI_{100i}$$

For a single axial scan at the z=0 location in a long phantom, a pencil chamber (100mm) can integrate the complete dose profile and give the dose to the region irradiated according to the formula below

$$CTDI_{100N} = \frac{D_{pencil}(z)}{nT} = \frac{1}{nT} \sum_{i=1}^{N} f_i(z) \Delta z_{ipencil}$$

We conducted a simple experiment to test this result. In the experiment, 2 body phantoms (30cm length) are used for measuring exposure from a Siemens Sensation 64 slice CT scanner. Scans were done at 120 kVp, 200 mA, 1 second and 24 x 1.2 mm collimation in axial mode. The approach was to take the pencil chamber (10cm active length) and place it in three regions as shown in figure 4 below. The result was summed and gives a value for the CTDI$_{100N}$ described above. We can call this the total dose ($D(z)$). The benefit of this approach is that it allows us to continue using the well-established CTDI$_{100}$ and the phantoms already available. Readings are taken at the peripheral and central locations and a weighted CTDI is obtained. The volume CTDI can then be calculated the usual way. This was validated with an approach where we used a farmer chamber and measured the exposure at multiple locations along the entire length of the phantom. This resulted in a dose within 5% of the approach using a pencil chamber in 3 locations. See table 3. The results of this experiment will be expanded to multiple scanners in a larger study in a future report. A summary is in the results section.

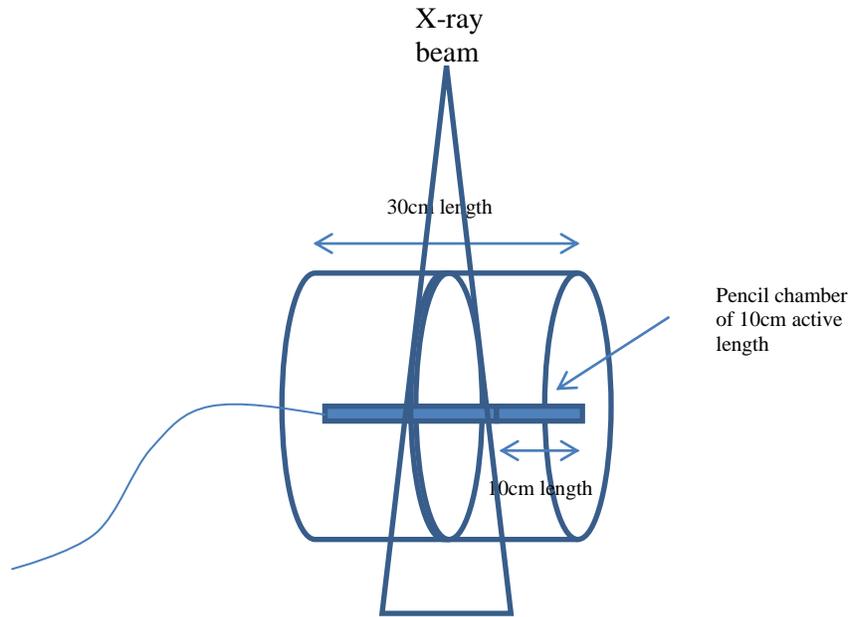

Figure. 4. Two body phantoms placed end to end. A pencil chamber is shown in 3 positions (a,b,c) where exposure readings can be collected. This was done for both center and peripheral regions of the phantoms. The resulting data can be used to calculate the total dose index $D_T(z) = CTDI_{300} = CTDI_{100a} + CTDI_{100b} + CTDI_{100c}$. Three exposures are taken at the central z=0 location of the phantom in axial mode for both center and peripheral axes.

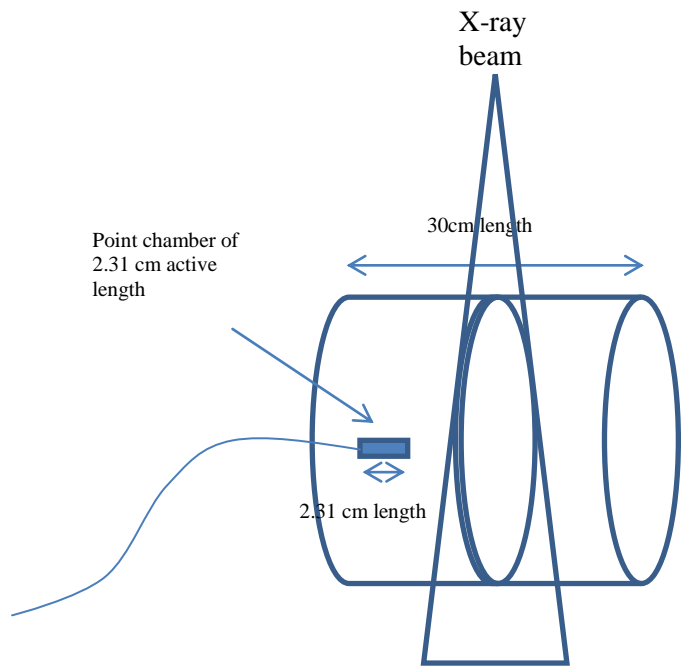

Fig 5. Two body phantoms placed end to end. A point chamber is shown in one position where an exposure reading can be collected for a shot taken at a location near the center of the phantom. This was done for both center and peripheral regions of the phantoms. The resulting data can be used to calculated the total dose $D_T(z) = CTDI_{300} = CTDI_{2.5a} + CTDI_{2.5b} + \ldots + CTDI_{2.5m}$. For the point chamber readings, thirteen (13) exposure readings were collected at 13 points along the entire length of the phantom. Shots were in axial mode at center and peripheral axes.

Let $f(z)$ be a function that describes the dose profile and fits the data. Then the area of the region under the graph of $f(z)$ is the exposure collected by a detector placed in that region under the graph. So for a 10cm pencil chamber, if the dose profile is 30 cm wide from one tail to the other, then we can sum the pencil chamber readings at each of three locations to get the total dose profile $D(z)$ below.

$$CTDI_{100N} = D_{pencil}(z) = \sum_{i=1}^{N=3} f(z_i)\Delta z_{ipencil}$$

Where the interval of integration is the length of the detector involved i.e., $\Delta z_i = 10cm$. It takes readings at three locations along the entire dose profile for both the center and peripheral axes to get the dose $D(z)$.

For a small or point chamber, the same formula can be written as below

$$CTDI_{23.1N} = D_{farmer}(z) = \sum_{i=1}^{N=13} f(z_i)\Delta z_{ifarmer}$$

In this case of a farmer chamber, the chamber is 2.31cm long so it will take about 13 readings at different locations along the entire dose profile for both center and peripheral readings to get the total dose $D(z)$. For the same dose profile, from above,

$$CTDI_{100N} = CTDI_{23.1N}$$

$$\sum_{i=1}^{N=3} f(z_{ipencil})\Delta z_{ipencil} = \sum_{i=1}^{N=13} f(z_{ifarmer})\Delta z_{ifarmer}$$

In general we can have a formula linking the two chambers as follows, where the integration is over each respective detector length. Note that the collimation (nT) set for comparing the two chambers cancels out in the formula below.

$$\sum_1^{N=3} \frac{KF_0 g}{nT}\left[\int_{-\frac{l}{2}}^{\frac{l}{2}}[y_p(x)y_s(z) + y_s(z)]\,dz\right] = \sum_1^{N=13} \frac{KF_0 g}{nT}\left[\int_{-\frac{l}{2}}^{\frac{l}{2}}[y_p(x)y_s(z) + y_s(z)]\,dz\right]$$

| 3 pencil chamber (100mm active length) readings at different locations =300mm | 13 Farmer chamber (23.1mm active length) readings at different locations =300.3mm |

This approach of using the pencil chamber or the farmer type chamber depends on using a large phantom that allows the scatter tails to converge to zero at large distances from the location of the scan. The scan is done without table movement and the detector is the only object moved from one end to the other. This approach is also analogous to the case of a long pencil chamber that is long enough to span the entire dose profile. An example will be the 30 cm long chamber used by Mori [11]. Several scenarios are depicted below. While the case described above for two body phantoms is from measurements of actual data (Table 3 below) from a Siemens Sensation 64 slice scanner, the scenarios described below are for a simulated dose profile and are used to illustrate the theory described above.

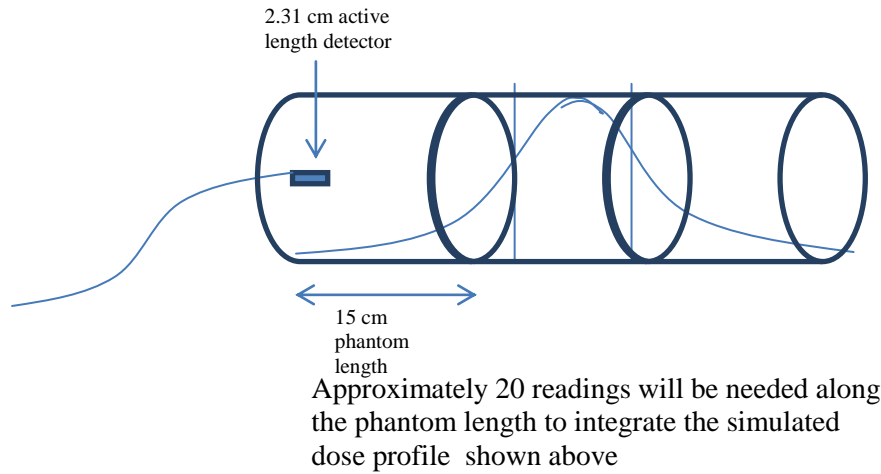

Approximately 20 readings will be needed along the phantom length to integrate the simulated dose profile  shown above

Fig6. A Farmer chamber of 2.31cm active length can be used in a 45cm phantom to integrate the entire dose profile from a single CT scan. This is done by moving the Farmer chamber while performing single scans in the same location until the entire dose profile is integrated. As usual center and peripheral axis measurements are taken to describe it as a $CTDI_n$. n is the number of times the dose profile is longer than that detector length. This is determined by phantom length required for convergence.

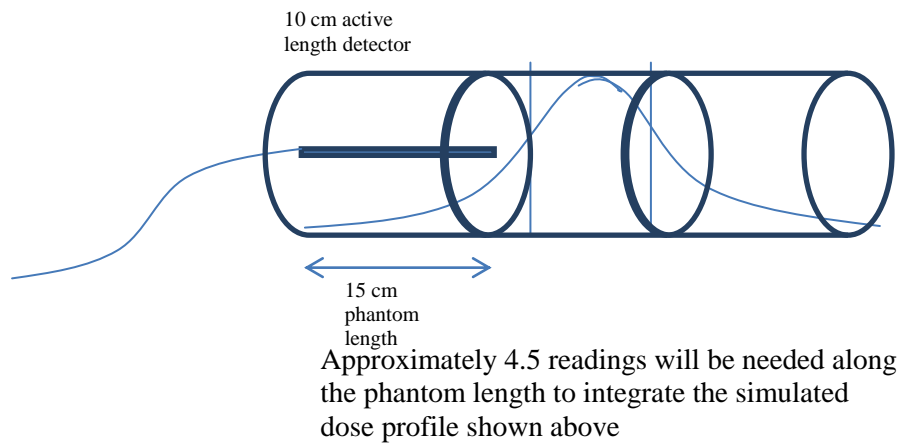

Approximately 4.5 readings will be needed along the phantom length to integrate the simulated dose profile shown above

Fig7. A pencil chamber of 10 cm active length can be used in a 45cm phantom to integrate the entire dose profile from a single CT scan. This is done by moving the pencil chamber while performing single scans in the same location until the entire dose profile is integrated. As usual center and peripheral axis measurements are taken to describe it as a CTDI$_n$. n is the number of times the dose profile is longer than that detector length. This is determined by phantom length required for convergence.

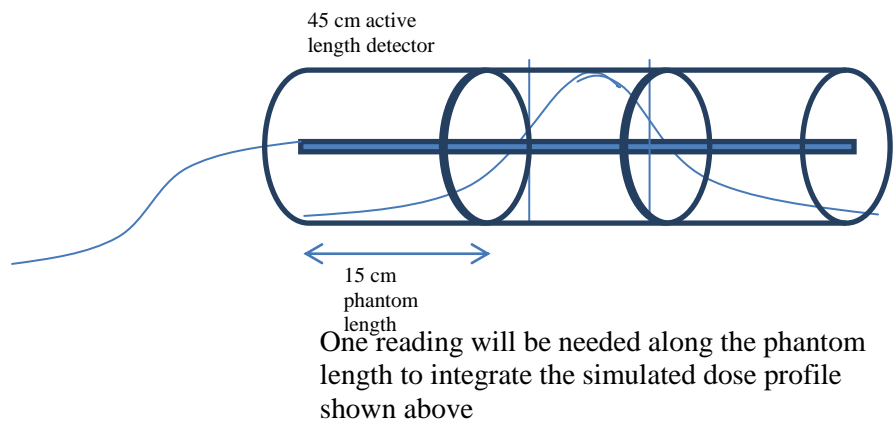

45 cm active
length detector

15 cm
phantom
length

One reading will be needed along the phantom
length to integrate the simulated dose profile
shown above

Fig8. A pencil chamber of 45cm active length can be used in a 45cm phantom to integrate the entire dose profile from a single CT scan. This is done by performing a single scan that integrates the entire dose profile. As usual center and peripheral axis measurements are taken to describe it as a CTDI$_n$. n is the number of times the dose profile is longer than that detector length. This is determined by phantom length required for convergence.

# RESULTS

## FITS OF THE MODEL TO THE O-ARM DATA

### Riccati's solution and the constructed function

Below are the fitted graphs for both the head and body phantoms using the complete model containing the scatter and constructed primary beam functions. Each of these graphs was fitted by using the total CT beam profile $f_T(z) = y_p(z)y_s(z) + y_s(z)$. The proportion of the primary and scatter components is determined by the regression analysis used for fitting the data. The equation representing the body dose profile is shown below. The primary and secondary parts are multiplied together and used for fitting the o-arm data in the collimated region. The region outside the collimation is fitted by adding in the secondary scatter component. The actual data does not extend to the tail regions since the phantom was not large enough to allow for convergence of the scatter tails as required. All graphs below are for data collected with three head or body phantoms and a point (Farmer) chamber.

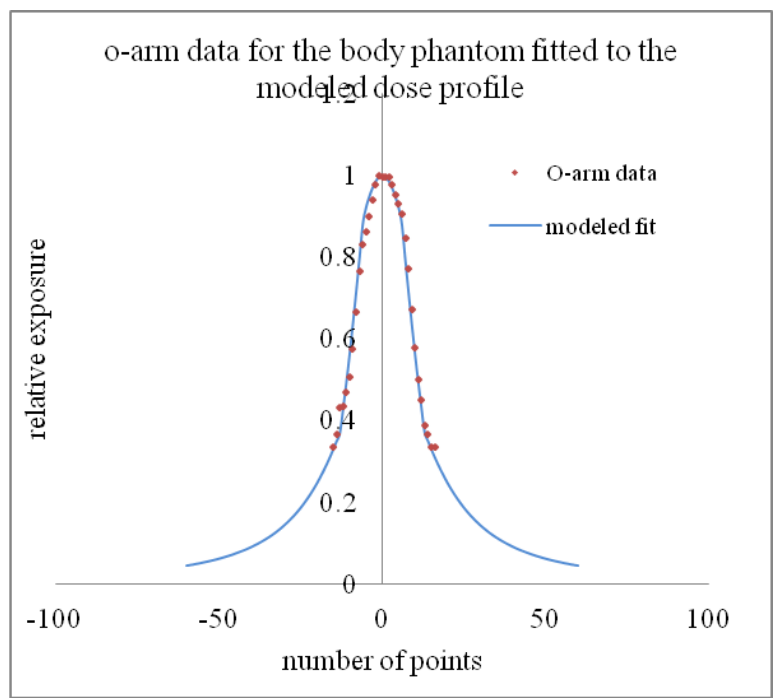

Fig9. The graph is normalized to a peak value of 1. The model shows an excellent fit to the data and is based on multiplying the solution to the Riccati equation with the constructed function for the collimated region and adding the secondary scatter component for the tails.

For the graph below the primary and secondary parts of the modeled functions are multiplied together and used for fitting the o-arm data for the head phantom in the collimated region. The scatter tail is fitted by adding in the secondary scatter component. The graph is shown below.

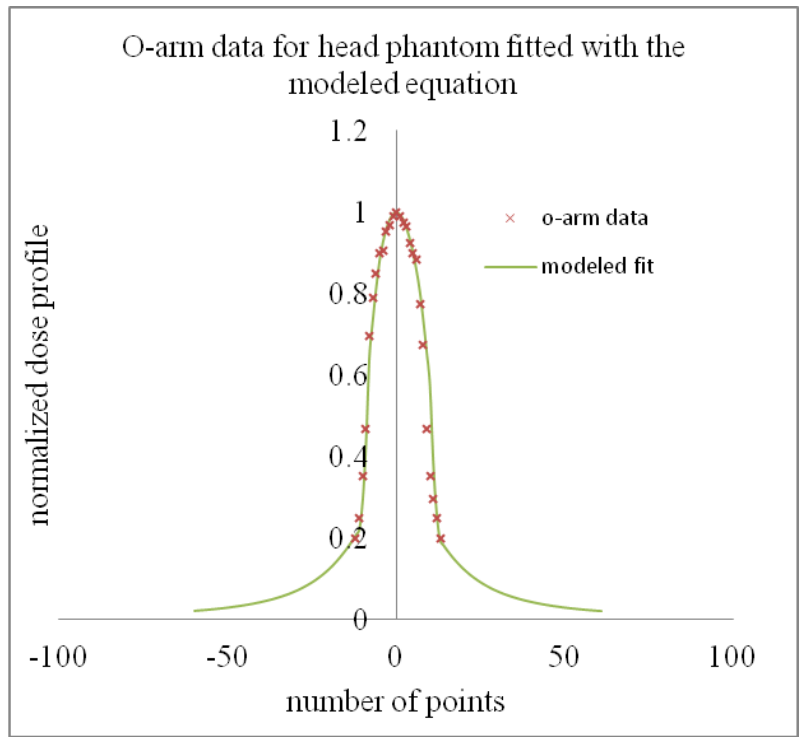

Fig. 10. Data for the o-arm was fitted by multiplying the primary and secondary parts of the modeled equations together to get the fits shown above for the collimated region. The scatter region is fit by adding in the secondary scatter component.

## Parameters used for the fits

The fits to the graphs above were improved by using a non-linear regression analysis. Parameters are in the table 1 below.

Table 1. Fit parameters used for fitting the dose profile model developed from the solution of the Riccati equation and a constructed function.

| | Fit parameters from non-linear regression analysis | | | | |
|---|---|---|---|---|---|
| | a | b | c | D | g |
| **Head phantom** | 3 | 9.5 | 0.041 | 8.5 | 0.99 |
| **Body Phantom** | 3.2 | 5.6 | 0.019 | 12 | 0.36 |

The first 3 fit parameters are for the scatter profile and the last 2 are for the primary beam profile.

# Riccati's solution and the trapezoidal function

The graphs below use the solution to Riccati's equation and the model of the trapezoidal function to produce the best fits to the o-arm data. The first graph is for the body phantom. Both functions were multiplied together for the collimated region and a scatter component added to make a total dose profile $y(z) = y_p(z)y_s(z) + y_s(z)$ for the head or body CT dose profiles.

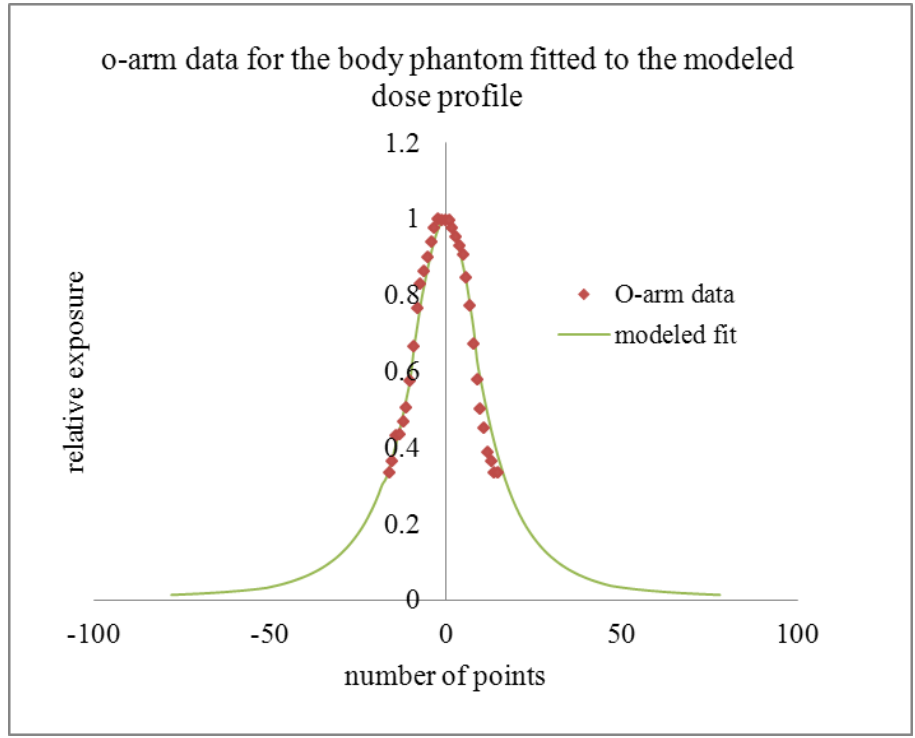

Fig 11. The graph is normalized to a peak value of 1. The model shows an excellent fit to the data and is based on multiplying the solution to the Riccati equation with the trapezoidal function for the collimated region and adding a secondary scatter component.

For the head phantom, the graph is shown below

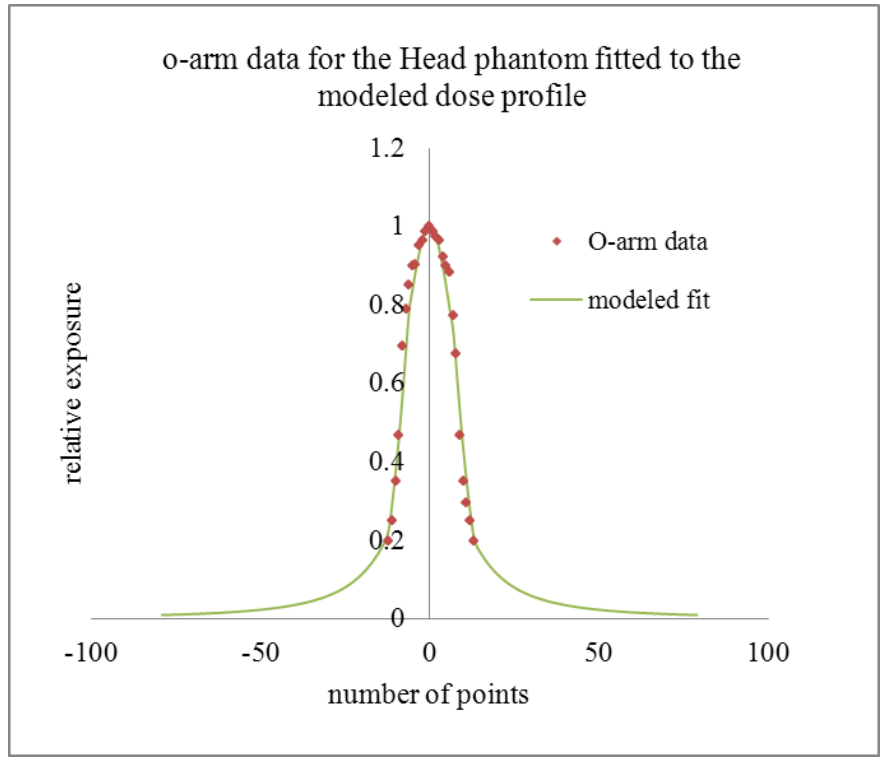

Fig 12 This o-arm data was fitted by multiplying the primary and secondary parts of the modeled equations together and then adding the secondary components to get the fits shown above. The equation is

$$f_T(z) = y_p(z)y_s(z) + y_s(z)$$

## Parameters used for the fits

The fits were improved by using a non-linear regression analysis. Parameters are in the table 2 below.

Table 2. Fit parameters used for fitting the dose profile model developed from the solution of the Riccati equation and a trapezoidal function.

|  | Fit parameters from non-linear regression analysis | | |
|---|---|---|---|
|  | a | b | C |
| **Head phantom** | 3 | 2.8 | 0.021 |
| **Body Phantom** | 3 | 6.4 | 0.036 |

# Gaussian function and the trapezoid function

The graph below is based on the trapezoid function and the Gaussian function to produce the best fits to the o-arm data. Only the head phantom is shown for this case. Both primary and scatter functions were multiplied together to fit the collimated region and the scatter component added to fit the tail regions. This results in a total dose profile $y(z) = y_p(z)y_s(z) + y_s(z)$.

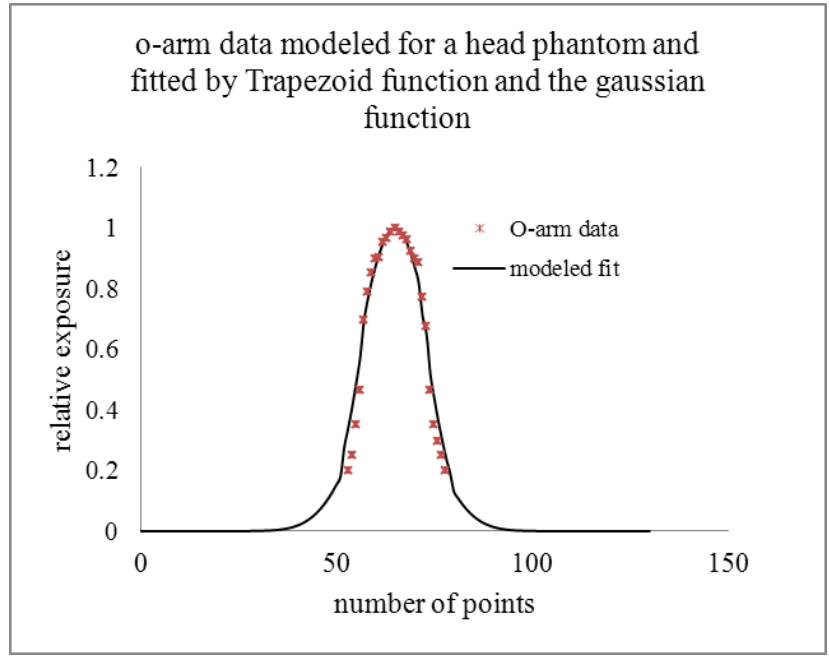

Fig 13. The graph is normalized to a peak of 1 for convenience. This case also shows data for the o-arm fitted to the model by multiplying the primary and secondary parts of the modeled equations together and adding the secondary scatter tails.

This graph uses a trapezoid function with an umbra and a penumbra region as well as the heel effect modeled in. It is multiplied with the Gaussian function to get the fit to the o-arm head data above in the collimated region and the scatter tails are added to fir the regions outside the collimation. In all the graphs shown above, the dose profile is fitted by the formula

$$f_T(z) = y_p(z)y_s(z) + y_s(z)$$

# PHANTOM DOSES MEASURED BY FARMER CHAMBER USING THE TG111 APPROACH COMPARED TO PENCIL CHAMBER READINGS

## Connection to CTDI

The results of the experiment described above are summarized here. Our approach collects the entire exposure as well as scatter tails, and actually gives a higher CTDI than is displayed on the scanner console (~72% more). The TG111 approach of using a farmer type chamber to measure the exposure in a large phantom at multiple locations gives a result that is within 5% of the reading obtained by using our approach of a pencil chamber in the same phantom, and measuring the exposure at multiple locations also. This approach was described and formulated above in section 2.2A.

The table 3 below shows the results of my approach compared to using the point chamber (TG111) approach. There is a ~%3 difference between the two approaches.

Table 3. Data was collected at central and peripheral positions using a farmer chamber and the 100 mm active length pencil chamber. The number of phantoms used for collection is included in the columns.

| Reported on console by machine | pencil chamber with 1 phantom | pencil chamber in center of 2 phantoms (1 reading) | Farmer (from one end to another of phantom 13 positions) | pencil(3 positions) |
| --- | --- | --- | --- | --- |
| 13.48mGy | 14.01mGy | 17.6mGy | 23.87mGy | 23.2mGy |

Table of dose indices measured using a pencil chamber and a farmer chamber in a body phantom. The first column denotes the CTDI displayed on the console.

# DISCUSSION

Although there are several functions that can be used to model dose profiles in CT, we chose to model scatter component of the dose profile as a solution to the Riccati equation. Comparing our solution to Riccati's equation to the other dose profile functions, we can conclude that the solution to Riccati's equation is most closely related to the Lorentzian function in form. Lorentzians were used earlier by Tsai ( 2003) and also Gagne (1989) to accurately fit the dose profiles in their work. The difference between this work and the work of Tsai and Gagne is that this work combines the solution to Riccati's equation with a primary function, either constructed or Trapezoidal, to get the total dose profile function. In the two works mentioned, the Lorentzian was used by itself to fit the dose profile function. This approach resulted in accurate fits to the narrow dose profiles in those works. The approach used in this paper of multiplying by the primary beam profile function before fitting will work for both narrow and wide beams in the collimated region. The scatter tail outside the collimated region is modeled by adding in the secondary scatter term according to the formula $f_T(z) = y_p(z)y_s(z) + y_s(z)$ . The key is to start with a primary beam profile with a width that is equal to the FWHM of the actual beam. This can be modeled into the trapezoidal function or a suitably constructed function as we have shown. Once the scatter function is also modeled with a FWHM equal to the actual beam FWHM, it gives accurate results each time for both narrow and wide beams alike in the collimated region, for both a flat panel system like the o-arm and can be used for MDCT scanners with wider beams. Outside the collimated region, the scatter tails can be modeled by adding the scatter component to the product of the primary and scatter. Compared to other works where a bi-exponential function or a Gaussian function was used, our approach is still consistent. Without multiplying by the primary beam function, and following the approach of matching the FWHM, these functions can fit some narrower dose profiles but are not as accurate for the wider profile of the cone beam system found on the o-arm. The shape of the primary beam can be considered to be a triangle for very narrow collimations. For larger collimations, the shape becomes more of a trapezoid. The use of a trapezoid also allows us to have flexibility in modeling the penumbra regions of the beam profile. Also the heel effect can be modeled easily in this case. In most of our models we did not include the heel effect except the case of the Gaussian and trapezoidal function, figure 13. Another approach we took involved the construction of a function based on the use of the analytic continuation of the Heaviside step function and a penumbra region constructed from linear functions. The constructed function also was combined with the Riccati solution successfully, figure 9 and 10. Another possibility is to think of the scatter dose profile as a Gaussian or Poisson distribution. If that is used, we can think of the scatter from the phantom as creating a distribution of photons that can be used in combination with the trapezoidal function to fit the dose profile in CT. This is shown in figure 13.

Our model is in contrast to other models that have been used in the past such as the Bi-exponential model of Shope et.al, and models based on the convolution of a unit rectangular function with a scatter function based on the use of the LSF (Dixon et.al, Boone). Our model provides a fit to the data from an O-arm flat panel system, a SCBCT system, and is flexible enough to be adapted to other MDCT systems as described above. This is done without the use of the convolution approach. One critique that we may have to accept is the lack of modeling of the heel effect in all cases. In our case, the heel effect did not introduce a significant improvement to the fits of the model to the data. Even for the case where it was modeled; the Gaussian and trapezoidal model case, the effect was not a significant improvement. Another criticism may be due to the use of the product of the primary and scatter component to the total dose profile in the collimated region. We believe this is valid since the scatter is a modulation of the primary beam by the phantom. There is still some work to be done to accurately and clearly define the various parameters in the models and to analyze the influence of each on the dose profiles and actual doses in CT scanning.

In fitting the model functions to the actual o-arm data, a nonlinear regression analysis was used. Due to the asymmetric nature of the actual data, fits on one arm are not as tight as one would expect for a perfectly symmetric data. This can be seen in figure 9. The asymmetry of the actual o-arm data was due to the the way the data was collected (as described in the text above as well as the presence of a heel effect). Note that the shape of the primary profile, through the width of the FWHM, plays a large role in the shape of the overall dose profile. The scatter profile modulates the primary profile in the collimated region and rounds the edges, thereby improving the fits. This is important since it demonstrates the accuracy of the developed mathematical model. One shortcoming may be the need for a more sophisticated fitting program that can provide a smoother fit to the data. Any improvements to the model can be demonstrated by the minimization of the squared difference between the actual and modeled data in the nonlinear regression model. Another area that may need improvement is the parameters used for the fits.

Also in part 2 of this paper, our goal was to develop a means by which the pencil chamber can be used to measure the exposure from scanners with wide flat panel detectors, such as the inter-operative o-arm. We later extended our theory to wider MDCT scanners. Modern CT scanners are increasingly produced with wider detectors. A 150 mm phantom and a 100mm active length pencil chamber will typically underestimate the dose in the phantom due to the scatter tails extending beyond the collection region of the chamber. Dixon proposed abandoning the pencil chamber in favor of a point chamber that can be used with a larger phantom. The point chamber will be used to measure the equilibrium dose in a large phantom, proposed by Dixon [8], that simulated an infinite scan length ([3]). This method was later put forth in the TG111 report [24]. We take a different approach in this paper. We demonstrated that a pencil chamber can still be used for measuring the dose if it is advanced by intervals equal to the length of the chamber if a large enough phantom is used to simulate convergence of the scatter tails of the dose profiles. We validated this approach both theoretically and experimentally. In particular, when the $D_{pencil}(z)$ integral is calculated at different endpoints to simulate the collection of exposure by an ion chamber, the results showed that if three equal areas were integrated the sum of these three areas is equal to the area obtained from just integrating over the same total endpoint. This is shown in table 3 in the results section. In addition, the results show that with two phantoms instead of one, more exposure is collected by the pencil chamber. Furthermore, with two phantoms, both pencil and farmer chamber collected more exposure than when one phantom was used. We expect that with more phantoms, the exposure collected should reach an equilibrium value. Several papers have described similar approaches to integrating the total dose profile by either moving the detector or moving the phantom while the detector is kept fixed. These papers have not shown the mathematical analysis shown here. It is important to note that if the detector is moved a distance equal to the length of the detector, the integration is contiguous as shown in C3. This is a similar situation compared to the case where the detector is kept fixed in a phantom and the phantom is moved. However, this scenario is trickier since the phantom distance moved at each scan has to match the detector length to avoid a pitch. This scenario is analyzed in C5. Our careful analysis on both scenarios shows how the detector can be used to integrate the dose profile either by moving it in a fixed phantom or by moving a phantom in which the detector is kept fixed.

Therefore, with the new theory presented here, we can now use the pencil chamber to measure the dose from flat panel CT detectors or wider MDCT scanners. We have proven that we can obtain an accurate exposure by summing the exposure from moving the pencil chamber at intervals equal to the active length of the chamber, to cover the entire phantom region and to collect the entire exposure $f_T(z)$ including long scatter tails. This is of course similar to two other scenarios. One involves using multiple pencil chambers each in a different position in the phantom to cover the entire beam profile, or second, a long pencil chamber of the same length as the sum of the active region of the multiple pencil chambers used individually. Our approach will be easier and more feasible logistically than using a point/farmer chamber and scanning as depicted in TG111. Our approach also gave results that were accurate to within 5% of the TG111 approach. These approaches described here are of course symmetrical to just keeping the chamber fixed at the center of the phantom and moving the entire phantom over multiple locations when exposures are taken as we have described above.

# CONCLUSION

Our paper addresses a very important and looming problem for dosimetry of MDCT scanners. We have shown that a mathematical model that uses a primary trapezoidal beam profile and a scatter component based on a nonlinear ODE such as the Riccati equation can give accurate fits to the dose profiles from a wide cone beam CT system. The mathematical approach is robust and logical and the model accurately describes observed data in the region containing actual data. We also propose a method of dosimetry for both an MDCT system and a wide cone beam CT. The dosimetry for any CT system with narrow or wide cone beam system can be done by using the current pencil chamber and multiple body or head CT phantoms to accurately measure exposure. The results of the mathematical theory as well as the proposed CTDI method have been shown to be accurate when compared to a current proposed method in TG-111 while being simpler to implement